\documentclass[useAMS,usenatbib,usegraphicx]{mn2e}
\usepackage{times, graphicx, setspace, amsmath, amssymb, latexsym, fancyhdr, caption, subfig, url}
\usepackage[toc]{appendix}

\title[A study of spatial correlations in PTA data]{A study of spatial correlations in pulsar timing array data}
\author
[C. Tiburzi et~al.]{C. Tiburzi$^{1,2,3,4}$\thanks{E-mail:ctiburzi@oa-cagliari.inaf.it}, G. Hobbs$^{5}$,  M. Kerr$^{5}$, W. A. Coles$^{6}$, S. Dai$^{5,7}$,  \newauthor R. N. Manchester$^{5}$, A. Possenti$^{1}$, R. M. Shannon$^{5,8}$,  X. P. You$^{9}$\\
\\
$^{1}$INAF - Osservatorio Astronomico di Cagliari, Via della Scienza, 09047 Selargius (CA), Italy\\ 
$^{2}$Dipartimento di Fisica, Universit\`a di Cagliari, Cittadella Universitaria 09042 Monserrato (CA), Italy \\
$^{3}$Max-Planck-Institut f\"{u}r Radioastronomie, Auf dem H\"{u}gel 69, 53121 Bonn, Germany\\
$^{4}$Fakult\"{a}t f\"{u}r Physik, Universit\"{a}t Bielefeld, Postfach 100131, 33501 Bielefeld, Germany\\
$^{5}$CSIRO Astronomy and Space Science, PO Box 76, Epping, NSW 1710, Australia \\
$^{6}$Electrical and Computer Engineering, University of California at San Diego, La Jolla, CA 92093-0407, USA\\
$^{7}$Department of Astronomy, School of Physics, Peking University, Beijing, 100871, China\\
$^{8}$Australia and International Centre for Radio Astronomy Research, Curtin University, Bentley WA 6102, Australia \\
$^{9}$School of Physical Science and Technology, Southwest University, Chongqing 400715, China\\
}

\begin{document}

\date{Accepted on 2015 September 9}
\pagerange{\pageref{firstpage}--\pageref{lastpage}} \pubyear{}

\maketitle
\label{firstpage}
 
\begin{abstract}
Pulsar timing array experiments search for phenomena that produce
angular correlations in the arrival times of signals from millisecond
pulsars. The primary goal is to detect an isotropic
and stochastic gravitational wave background. We use simulated data to
show that this search can be affected by the presence of other
spatially correlated noise, such as errors in the reference time
standard, errors in the planetary ephemeris, the solar wind and instrumentation issues. All these effects can induce significant false
detections of gravitational waves. We test mitigation routines to
account for clock errors, ephemeris errors and the solar wind. We
demonstrate that it is non-trivial to find an effective mitigation
routine for the planetary ephemeris and emphasise that other
spatially correlated signals may be present in the data.
\end{abstract}


\section{Introduction}\label{Sec:intro}

The pulsar timing method provides a way to study the physics and
astrometry of individual pulsars, the properties of the intervening
interstellar medium and to search for un-modeled phenomena. Pulsar
times-of-arrival (ToAs) are first measured at radio observatories and
then converted to the reference frame of the solar-system barycenter
(SSB). The extreme rotational stability of pulsars enable these
barycentric ToAs to be predicted using a simple model (known as the
``pulsar ephemeris'') for the pulsar's rotational and orbital
motion. Differences between the measured and predicted ToAs are
referred to as ``timing residuals''. Features in the timing
residuals (which imply a temporal correlation between the ToAs of an
individual pulsar) indicate that the model with which we are
describing the pulsar is not complete. It may not account for some
events, such as glitches (e.g., \citealt{ww12}), variations in the
interstellar medium (e.g., \citealt{kcs13}) or intrinsic instabilities
in the pulsar rotation (e.g., \citealt{sc10}). The power spectrum of
such residuals is often ``red'', that is, characterized by an excess
of power at lower fluctuation frequencies. In contrast, ``white" power
spectra have statistically the same amount of power for each
frequency. An iterative process is followed to improve the pulsar
model in order to minimize the residuals. Numerous studies have
already been carried out using this technique, including tests of
theories of gravity (e.g., \citealt{ksm+06}), analyses of the
Galactic pulsar population (e.g., \citealt{lfl+06}) and the
interstellar medium (e.g., Keith et al. 2013).

Pulsar Timing Array (PTA) projects\footnote{There are three
  collaborations in the world that are leading PTA experiments: the
  Parkes PTA (PPTA) in Australia \citep{mh13}, the European PTA (EPTA)
  in Europe \citep{kc13} and the North America Nanohertz Observatory
  for gravitational waves (NANOgrav) in North America
  \citep{m13}. These teams have joined together to establish the
  International PTA (IPTA; \citealt{ha10,man13}).} are employing this
technique to study phenomena that simultaneously affect multiple
pulsars. A main goal of PTA experiments is the direct detection of an
isotropic and stochastic gravitational wave background (GWB; e.g.,
\citealt{hd83}) generated by the superposition of the
gravitational-wave emission of coalescing super-massive black-hole
binaries \citep{ses13}. Other noise processes that simultaneously
affect many pulsars are, for example, irregularities in terrestrial
time standards \citep{hc12} and poorly determined solar system
ephemerides \citep{ch10}.

In all of the cases mentioned above, the pulsar ToAs will be spatially
and temporally correlated.  PTA projects seek to measure the spatial
correlation $C(\theta_{ij})$ between a pair of pulsars (labeled $i$
and $j$) separated by an angle $\theta_{ij}$. The $C(\theta_{ij})$ coefficients are
analyzed to identify the physical phenomenon that leads to the
correlation. In particular, a GWB would induce a specific shape in the spatial correlations (see Figure~\ref{Fig:hdcurve}), called the ``Hellings and Downs curve'' \citep{hd83}. If the Hellings and Downs curve is significantly detected, then a GWB detection will be claimed.

To date, no GWB detection has been made. Because the first direct
detection of GWs will be of enormous astrophysical interest, the
chance of false detections must be well understood. After a detection,
the first step will be to determine an unbiased estimate of the
properties of that background (such as its amplitude). It is therefore
fundamental to verify whether any other physical effects could lead to
an angular correlation whose cause could be misidentified as a GWB.

In this paper, we study the impact of various noise processes that would correlate between different pulsars on the GWB search in PTA data. In particular, we study
simplified versions of the signals that we expect from realistic
errors in the clock time standard and in the planetary ephemeris. We
also discuss a set of mitigation routines to remove the effect of
these noise processes before providing an estimate of real errors in the
clock and planetary ephemeris. We consider other possible correlated
noise processes that might be hidden in the data.

To carry out our research, we use simulated data sets. In
Section~\ref{Sec:ang_correlations} we outline the properties of the
noise processes that we simulate. In Section~\ref{Sec:method} we
describe how we realize these simulations and their processing. In
Section~\ref{Sec:res_disc} we present and discuss our results, and
draw our conclusions in Section~\ref{Sec:conclusions}.

\section{Correlated signals in pulsar data}\label{Sec:ang_correlations}

PTA pulsars are generally widely separated ($\gg 1$\,degree) on the
sky.  Noise induced by the interstellar medium and the intrinsic
timing noise (unexplained low-frequency noise in the timing residuals of a single pulsar) of each individual pulsar is expected to be uncorrelated
between pulsar pairs:
 
 \begin{equation}
C(\theta_{\rm ij}) = 0.
\end{equation}

Pulse ToAs are referred to an observatory time standard. However,
observatory clocks are not perfect and so the ToAs are converted to a
realization of Terrestrial Time (TT). This conversion is carried out
with a precision usually better than $\sim$1\,ns; see \citet{hc12}.
Two main realizations of TT are commonly used.  Terrestrial Time as
realized by International Atomic Time (TAI) is a quasi-real-time time
standard.  This is subsequently updated to produce the world's best
atomic time standard, the Terrestrial Time as realized by the Bureau
International des Poids et Mesures (BIPM). Any errors in the
Terrestrial Time standard will induce the same timing residuals in all
pulsars:

\begin{equation}
C(\theta_{\rm ij}) = 1,
\end{equation}\label{Eq:monopole}
or, the correlation between pulsars affected by an error in the clock
time standard is monopolar. 

The pulsar timing procedure also relies upon knowledge of the position
of the SSB with respect to the observatory. The position of the
observatory with respect to the center of the Earth is usually
precisely known. In this case, we only need to consider possible
errors in the solar system ephemeris used to convert pulse ToAs from
the Earth's center to the SSB. The Jet Propulsion Laboratory (JPL)
series of ephemerides are often used for this conversion. In this
paper we make use of the ephemerides
DE414\footnote{\url{ftp://ssd.jpl.nasa.gov/pub/eph/planets/ioms/de414.iom.ps}}
and
DE421\footnote{\url{http://ipnpr.jpl.nasa.gov/progress_report/42-178/178C.ps}}.
The effect of an error in the planetary ephemeris on the pulsar timing
residuals, $r$, is:

\begin{equation}\label{Eq:res_pe}
r_{\rm i}(t) = \frac{1}{c} {\bf e}(t)\cdot {\bf \hat{k}}_{\rm i}
\end{equation}
where $c$ is the vacuum speed of light, ${\bf e}$ is the
time-dependent error in position of the SSB with respect to
the observatory and ${\bf \hat{k}}_{\rm i}$ is a unit vector pointed toward
pulsar $i$. Instantaneously, this is a dipolar effect. If the vector ${\bf e}$ is sufficiently independent of the ${\bf \hat{k}}_{\rm i}$ and ${\bf \hat{k}}_{\rm j}$ vectors, (as discussed in Appendix~\ref{App:a1}) then the $C(\theta_{ij})$ would assume a $\cos(\theta_{ij})$ angular dependence.

The ToA delays induced by the passage of a GW includes the effect of
the GW passing the pulsar (often known as the ``pulsar term'') and
also the effect of the GW passing the Earth (the ``Earth
term''). Recent work \citep{svv09,rw15} suggests that the most likely
detectable signal will be a stochastic and isotropic background of GWs
(GWB) caused by a large number of supermassive, binary black-hole
systems. The ``pulsar term'' will lead to red, uncorrelated noise in
the pulsar timing residuals.  \citet{hd83} showed that the
correlations induced by the ``Earth term'' would leave a well-defined
signature in the angular correlation between the timing residuals of
different pulsars. The signature, known as the ``Hellings \& Downs
curve'', is shown as the continuous line in Figure~\ref{Fig:hdcurve}
and is defined by:

\begin{equation}\label{Eq:hdcurve}
C(\theta_{\rm ij}) = \frac{3}{2}x\log(x) - \frac{x}{4} + \frac{1}{2}
\end{equation}

\noindent where $x = [1 - \cos(\theta_{\rm ij})]/2$. The power
spectrum for a GWB may be described by a power law:

\begin{equation}\label{Eq:ps_gwb}
 P_{\rm GWB}(f) = \frac{\rm
   A^2}{12\pi^2}\left(\frac{f}{f_{\mathrm{yr}}}\right)^{2\alpha -3}
\end{equation}

\noindent where A is the GWB amplitude for a frequency $f =
f_{\mathrm{yr}} = (1 \mathrm{yr})^{-1}$, $\alpha$ sets the power-law
slope, which is predicted to be $-2/3$ for an isotropic and
stochastic GWB \citep{p01}. Current upper bounds on the amplitude of
the GWB indicate that A is smaller than 10$^{-15}$ (Shannon et al.,
submitted).

The search for the GWB is based on determining the correlation between
the timing residuals for each pair of pulsars in a given PTA. The
subsequent steps of the search depend on the adopted statistical
approach. The PPTA have generally made us of a frequentist method (e.g., Yardley et al. 2011\nocite{yc11}, hereafter Y11) to determine whether or
not the correlations take the form of the Hellings \& Downs curve. If
they do, then a detection of the GWB will be claimed. We note that
this functional form will never be perfectly matched in
practice. First, the Hellings \& Downs curve is not obtained through
independent measurements of the angular covariance.  For a given PTA,
only a finite number of pulsar pairs exists and the measured
correlations will not be independent as a given pulsar will contribute
to multiple pairs. The Hellings \& Downs curve is an ensemble average. 
For our Universe, the positions and properties
of the black-hole binaries along with the effect of the GWB passing
each pulsar will lead to noise on the expected curve (see e.g.,
\citealt{rwh12}). Various algorithms, both frequentist and Bayesian,
have been developed to search for the signature of the Hellings \&
Downs curve \citep{yc11, vhl11, df13, lt15} and have been applied to
actual data sets.

If a non-GWB noise process that produces spatially correlated timing
residuals is well understood then it will be possible to develop
mitigation routines or to include such noise as part of GWB detection
algorithms.  We have already considered two signals (clock and
ephemeris errors) that are extensively discussed by PTA research
teams, but note that many other such processes may be present. These
include instrumental artifacts, solar wind effects, polarization
calibration errors, uncertainties in the Earth-orientation-parameters,
the ionosphere, the troposphere, individual gravitational wave
sources and many other possibilities.  For this paper, we have chosen
to study instrumental delays and the solar wind as representative
examples.

The configuration of the observational set-up for receivers and
signal-processing systems is optimized on the basis of the pulse
period and dispersion measure (DM) for each pulsar. Therefore, it may be that different
pulsars are characterized by different observational set-ups. For all
PTAs, both receivers and signal-processing systems have evolved over
the data spans (see, e.g., \citealt{mh13} for details). Such updates
can introduce time offsets into the timing residuals that are
dependent upon the specific configuration used. Although PTA teams
attempt to compute them, these measurements have an uncertainty, and
thus some small offsets remain. Therefore, the pulsar data sets may
include an offset that is identical for pulsar pairs with the same
observational configuration, but has a different amplitude for pulsar
pairs with different set-ups.

Fluctuations in the plasma density of the solar wind and the ionosphere will cause correlated noise in the timing residuals of different pulsars if the DM corrections are not made on a sufficiently fine time scale. As the the solar wind variations yield the most severe impact between these two effects, we have included them in our simulations. The solar wind variations depend on the angular distance between the pulsar and the Sun and the solar latitude of the point where the line of sight is closest to the Sun \citep{yhc+07a}. The size of this effect will change during the solar cycle (see e.g., \citealt{ych12}).  The \textsc{tempo2} software attempts to account
for the solar wind using a non-time-dependent and spherically
symmetric model (see e.g., \citealt{ehm06}). It is often assumed that
any remnant signal could be absorbed into standard measurements of DM
variations for each pulsar.  However, variations in the solar wind may
occur on time scales faster than the typical smoothing time for
dispersion measure correction (see e.g., \citealt{kcs13}), or be in
disagreement with some proposed models for DM fluctuations
(e.g., \citealt{lbj+14}).

\section{Method}\label{Sec:method}

\begin{table*}
 \footnotesize
\centering
\caption{Some parameters of the PPTA pulsar sample simulated in this study.}\label{Tab:ppta_msp}
\begin{tabular}{c c c c c}
\hline
\textbf{PSR name}&\textbf{Spin period}&\textbf{RA}        &\textbf{Dec} & \textbf{Ecliptic latitude}\\
                 &\textbf{[ms]}       &\textbf{[hh:mm:ss]}&\textbf{[dd:mm:ss]} & \textbf{[deg]}     \\
\hline
J0437$-$4715 &	5.757  & 04:37:15.8 &  $-$47:15:08.6 & $-$67.9 \\
J0613$-$0200 &	3.062  & 06:13:43.9 &  $-$02:00:47.1 & $-$25.4 \\
J0711$-$6830 &	5.491  & 07:11:54.2 &  $-$68:30:47.5 & $-$82.9 \\
J1022+1001 &	16.453 & 10:22:58.0 &  +10:01:53.2 &  $-$0.1 \\
J1024$-$0719 &	5.162  & 10:24:38.6 &  $-$07:19:19.1 & $-$16.0 \\
& & & & \\
J1045$-$4509 &	7.474  & 10:45:50.1 &  $-$45:09:54.1 & $-$47.7 \\
J1600$-$3053 &	3.598  & 16:00:51.9 &  $-$30:53:49.3 & $-$10.1 \\
J1603$-$7202 &	14.842 & 16:03:35.6 &  $-$72:02:32.7 & $-$50.0 \\
J1643$-$1224 &	4.622  & 16:43:38.1 &  $-$12:24:58.7 & +9.8 \\
J1713+0747 &	4.57   & 17:13:49.5 &  +07:47:37.5   &  +30.7 \\
& & & &\\
J1730$-$2304 &	8.123  & 17:30:21.6 &  $-$23:04:31.2 &  +0.2 \\
J1732$-$5049 &	5.313  & 17:32:47.7 &  $-$50:49:00.1 & $-$27.5 \\
J1744$-$1134 &	4.075  & 17:44:29.4 &  $-$11:34:54.6 &  +11.8 \\
J1857+0943 &	5.362  & 18:57:36.3 &  +09:43:17.3 &  +32.3 \\
J1909$-$3744 &	2.947  & 19:09:47.4 &  $-$37:44:14.3 & $-$15.2 \\
& & & &\\
J1939+2134 &	1.558 & 19:39:38.5 &  +21:34:59.1 &  +42.3 \\
J2124$-$3358 &	4.931 & 21:24:43.8 &  $-$33:58:44.6 & $-$17.8 \\
J2129$-$5721 &	3.726 & 21:29:22.7 &  $-$57:21:14.1 & $-$39.9 \\
J2145$-$0750 &	16.052 & 21:45:50.4 &  $-$07:50:18.4 &  +5.3 \\
J2241$-$5236 &  2.187 & 22:41:42.0 &  $-$52:36:36.2 & $-$40.4 \\
\hline                   
\end{tabular}
\end{table*}

\begin{figure}
\centering
 \includegraphics[scale=0.45,trim=0.5cm 0 0 0]{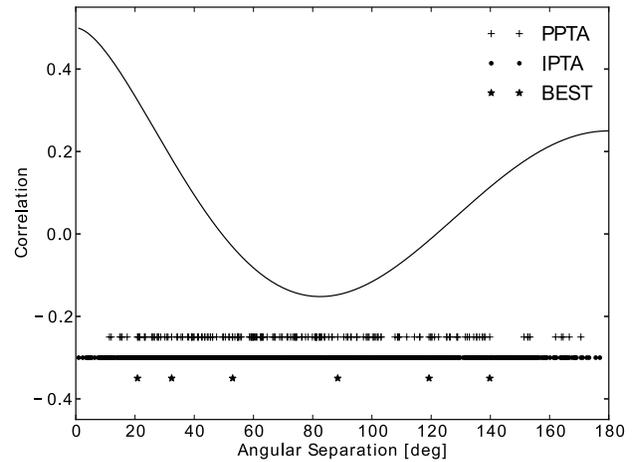}
\caption{The continuous line represents the expected Hellings \& Downs
  curve. Crosses, dots and stars show the angular coverage offered by,
  respectively, the PPTA pulsars, the IPTA pulsars (see Table
  \ref{Tab:ppta_msp}) and the four best PPTA (and IPTA) pulsars:
  PSRs~J0437$-$4715, J1713+0747, J1744$-$1134 and
  J1909$-$3744.}\label{Fig:hdcurve}
\end{figure} 

In order to study the effect of correlated noise processes on the GWB
search in PTA data we 1) simulate data sets in which we artificially
inject signals that might induce spatial correlations between pulsar
time series and 2) process each simulation as if only a GWB was
present in the data.  PTA data sets are subject to various
complexities: different pulsars may have different data spans, the
precision with which the ToAs can be determined is affected by the
flux density of the pulsar and interstellar scintillation and the
observational sampling is non-uniform.

In this paper, we have chosen to simulate much simpler (albeit
unrealistic) data sets, with regular sampling and equal uncertainties
and data spans. With such data sets, our software is faster and many
of the results can be understood analytically. We also note that if an
analysis of these data sets can lead to misidentifications of a GWB,
then the additional effects evident in actual data may provide an
even higher chance of false detections.

We simulate data sets for 20 of the millisecond pulsars (MSPs)
observed by the PPTA (listed in Table \ref{Tab:ppta_msp}). The
coverage of the Hellings \& Downs curve offered by these pulsars is
shown by the cross symbols in Figure~\ref{Fig:hdcurve}.  We note that
the closest pulsar pair is PSR~J2129$-$5721 - PSR~J2241$-$5236 with an
angular distance of $11.36$\,degrees, and the most widely separated is
PSR~J1022$+$1001 - PSR~J2145$-$0750 with an angular distance of
$170.57$\,degrees. Only nine pulsar pairs have angular separations
wider than $140$\,degrees. The angular coverage given by the four most precise timers of the three PTAs (J0437$-$4715, J1713$+$0747, J1744$-$1134 and J1909$-$3744) is also shown in Figure~\ref{Fig:hdcurve} (star symbols), and ranges from 20 to 140 degrees.

We simulate the ToAs using the \textsc{ptaSimulate} software
package. This software is based on the simulation routines developed
for \textsc{tempo2}.  Initially, the software simulates the observing
times.  In our case, we simply assume that each of the 20 PPTA pulsars
are observed from MJD 48000 to 53000 (a span of 5000\,days,
13.7\,years), with a cadence of 14\,days.  The software then uses the
deterministic pulsar timing models (obtained from the ATNF Pulsar
Catalogue\footnote{\url{http://www.atnf.csiro.au/research/pulsar/psrcat/}})
for each of the pulsars to form idealized site arrival times based on
the observation times (see \citealt{hj09} for details). These timing
models are referred to the JPL ephemeris DE421 and the time standard
TT(BIPM2013). Later, the software creates 1000 realizations of the various
noise processes, which we add to the arrival times together with 100\,ns of white Gaussian noise to
represent the radiometer noise level. 

We first simulate 1000 realizations of time series affected by an
isotropic, stochastic GWB, S$_{\rm gwb}$ with an amplitude of 10$^{-15}$. This was achieved by
assuming 1000 individual GW sources isotropically distributed on the
sky with amplitudes and frequencies chosen such that the resulting
power spectrum of the background is consistent with
Equation~\ref{Eq:ps_gwb} \citep{hj09}. S$_{\rm gwb}$ is our reference
simulation throughout this paper.

We then study the likelihood that each specific noise process can be
mis-identified as a GWB, by individually adding these signals to the
simulated arrival times described above.  We do not know the exact levels,
nor the spectral shapes that such signals will assume in real data
sets. For this reason, we choose the following approach. We first determine whether noise processes with
the same spectrum and amplitude of the GWB in S$_{\rm gwb}$ can lead
to false detections. The red noise processes given below were
therefore all simulated to have spectra consistent with that of a GWB
with an amplitude A = 10$^{-15}$:

\begin{itemize}
\item \textbf{S$_{\rm utn}$}, spatially uncorrelated timing noise.  A
  scaling factor, which converts the real and imaginary parts of a
  Fourier transform to the required power spectrum, is determined.
  The real and imaginary parts are randomized by multiplying by a
  Gaussian random value. The corresponding time series is obtained by
  carrying out a complex-to-real Fourier transform;
\item \textbf{S$_{\rm clk}$}, stochastic, monopolar clock-like
  signal. Its characteristics are based on our expectation of an error
  in the clock time standard. For each realization, we derive a single
  time series as for S$_{\rm utn}$ and assume that this noise process
  affects all the pulsars;
\item \textbf{S$_{\rm eph}$}, stochastic, dipolar ephemeris-like
  signal. Its characteristics are based on our expectation of an error
  in the planetary ephemeris. For each realization, we simulate three
  time series as for S$_{\rm utn}$.  We assume that these three time
  series represent the time series of errors in the three spatial
  components of ${\bf e}$.  The timing residuals for a given pulsar
  are then obtained using Equation~3.
\end{itemize}

We also simulate other four noise processes that are not at the same
level nor have the same spectral shape as the simulated GWB:

\begin{itemize}
 \item \textbf{S$_{\rm tt}$}, we simulate time series based on the
   TT(BIPM2013), and we then process the time series by using
   TT(TAI). We thus simulate the effects of a clock error
   corresponding to the difference between TT(BIPM2013) and TT(TAI);
 \item \textbf{S$_{\rm de}$}, we simulate time series based on the
   DE421, and we then process the time series by using DE414. We thus
   simulate the effects of an ephemeris error corresponding to the
   difference between DE421 and DE414;
 \item \textbf{S$_{\rm ie}$}, we divide the pulsars of our sample in three groups (consisting of 7, 7 and 6 pulsars respectively).  We assume that each group was observed with a different signal-processing system.  We assume that the systems are modified six times over the total data span\footnote{The actual PPTA instruments were upgraded with approximately this cadence.}. This introduces time delays in the ToAs. We assume that the epochs of the modifications are the same for the three systems, but that the amplitudes of the time offsets are different. We also assume that these offsets vary for the same signal-processing system from one epoch to the other. The amplitude of the offsets (for each epoch and system) was randomly drawn from a Gaussian population a standard deviation of 20\,ns. We also test the effects for a standard deviation of 200\,ns;
 \item \textbf{S$_{\rm sw}$}, we include the expected effects of the
   solar wind by modeling the solar corona using observations from
   the Wilcox Solar Observatory\footnote{http://wso.stanford.edu/}
   when calculating the idealized site arrival times. This process is based on the procedure first described by \citet{yhc+07a}. We subsequently
   processed the pulsar observations assuming that the solar wind is
   absent.
\end{itemize}

These simulations are listed in Table~\ref{Tab:simulations}. We note that the GWB simulations require knowledge of the pulsar distances. In our case, the code computes an approximate distance based on the DM measurement. We also note that S$_{\rm tt}$, S$_{\rm de}$ and S$_{\rm sw}$ are all deterministic simulations, and that the only change between the 1000 iterations is the realization of the white noise for each pulsar.

 \begin{table}
 \small
 \centering
 \caption{Summary of the simulations. All the simulations include 1000 iterations of the timing residuals of 20 pulsars with a white noise level of 100\,ns, each spanning 5000 days and sampled at 14 day intervals. The simulations that assume a GWB spectrum have a spectral amplitude of 1$\times 10^{-15}$.}\label{Tab:simulations}
 \begin{tabular}{l l c }
 \hline
  \textbf{Tag} & \textbf{Simulated effect} & \textbf{GWB-like} \\
   &  &  \textbf{spectrum}\\
  \hline
  S$_{\rm gwb}$ & GWB & Y \\
  S$_{\rm utn}$ & Uncorrelated red noise & Y \\
  S$_{\rm clk}$ & Stochastic clock-like errors & Y \\
  S$_{\rm eph}$ & Stochastic ephemeris-like errors & Y \\
  \hline
  S$_{\rm tt}$  & Difference between TT(BIPM2013) and TT(TAI) & N \\
  S$_{\rm de}$  & Difference between DE421 and DE414 & N \\
  \hline
  S$_{\rm ie}$ & Instrumental errors & N \\
  S$_{\rm sw}$ & Solar wind & N \\
  \hline
  \end{tabular}
 \end{table}

\subsection{Simulation processing and computation of the angular covariance}

\begin{figure*}
 \centering
 \includegraphics[scale=0.55, trim = 0 0 0 0]{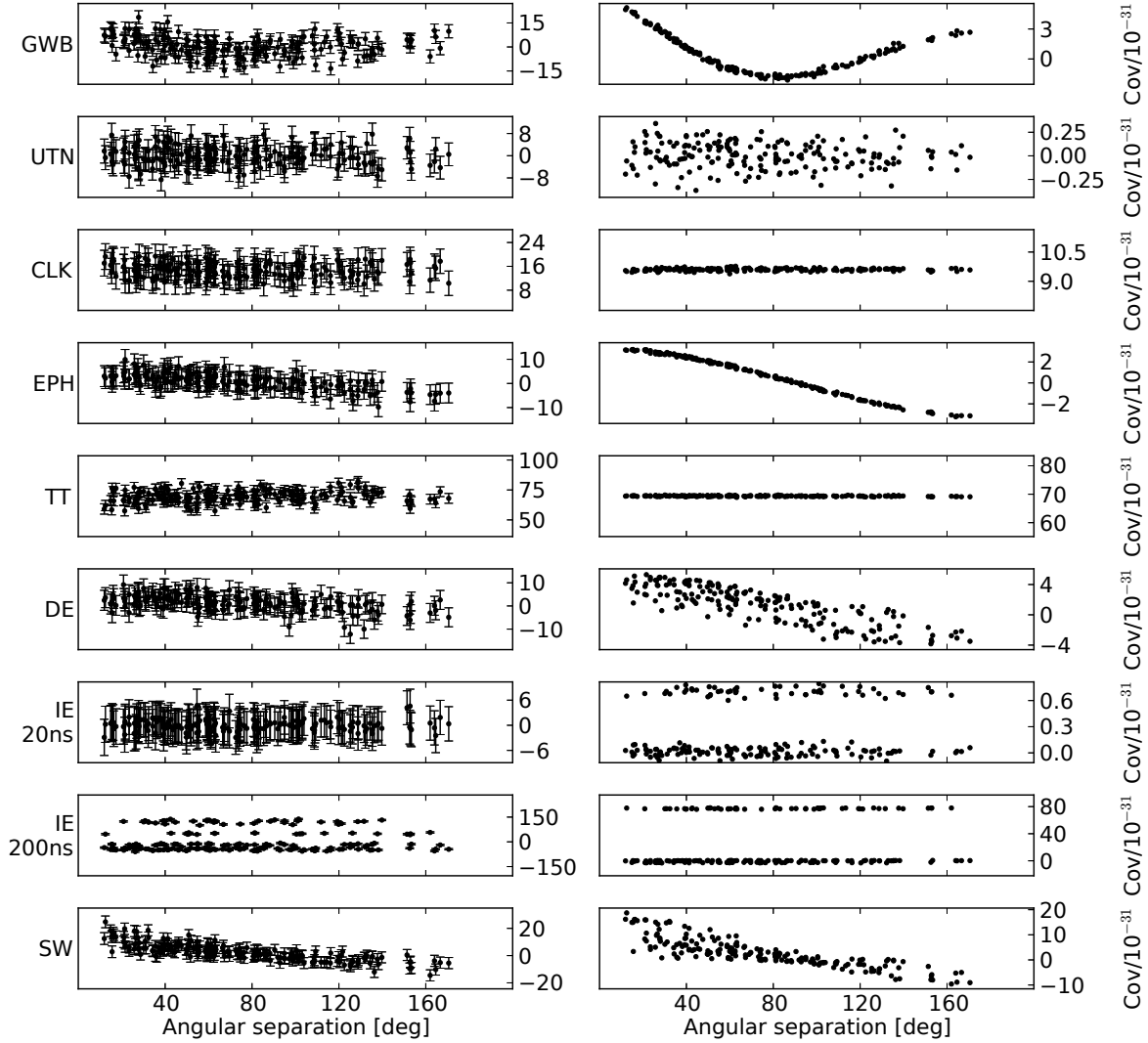}
\caption{Angular covariances computed the simulations described in Section~\ref{Sec:method}. From top to bottom: S$_{\rm gwb}$, S$_{\rm utn}$, S$_{\rm clk}$, S$_{\rm eph}$, S$_{\rm tt}$, S$_{\rm de}$, S$_{\rm ie}$ at 20\,ns, S$_{\rm ie}$ at 200\,ns, S$_{\rm sw}$. In the left panels is shown the result for an individual realization, in the right panels is shown the average over 1000 realizations.}\label{Fig:all_hd}
\end{figure*}

Our simulation software provides a set of pulsar timing models and
arrival time files that can be processed in an identical manner to
actual pulsar data. These timing models have already been fitted for
changes in the pulse frequency and its time derivative for each pulsar (and,
in the case of S$_{\rm de}$, we also fit for the pulsar position). We
wish to extract from these simulated data sets the correlated noise
that could lead to a false detection of a GWB.  Numerous GWB detection
codes exist \citep{yc11, vhl11, df13, lt15}. We chose to use an
updated version of the algorithms developed by Y11 because 1) it has been implemented in the
\textsc{tempo2} public plugin \textsc{detectGWB}, 2) the code runs
quickly, 3) it produces a large amount of useful diagnostic output and
4) it is easy for us to implement the mitigation methods that we
describe below.  The Y11 code first calculates the covariances between
the timing residuals for each pulsar pair.  It then obtains a value
for the squared GWB amplitude, $\hat{\rm A}^2$, by fitting the Hellings \& Downs
curve to these covariance estimates.  We therefore obtain an estimate of the GWB amplitude for each of our 1000
realizations.

We have updated the Y11 algorithm as follows. We linearly interpolate a grid of uniformly spaced values onto the pulsar timing residuals\footnote{The Y11 algorithm can be applied to real, unevenly sampled observations. However, this step both reduces the number of observations that need to be processed and ensures that they are on a convenient evenly spaced grid.}. The obtained grid of values is
constrained so that the corresponding rotational frequency and its
derivative are zero.  We then select the grid points that are in
common between a given pulsar pair and form cross power spectra using
a red noise model and the Cholesky fitting routines (see
\citealt{ch11}).  This allows us to calculate power spectra in the
presence of steep red noise. The noise model that we assume throughout
is given by Equation~\ref{Eq:ps_gwb} with A = 10$^{-15}$. The Y11
algorithm also requires an initial guess for the GWB amplitude.  For
this guess we also chose A = 10$^{-15}$. This implies that the
algorithm will be optimal for the case in which the GWB is present and
allows us to evaluate the occurrence of false detections in the other
data sets.
 
The output from the Y11 algorithm includes both an evaluation of A$^2$ ($\hat{\rm A}^2$), a 1-$\sigma$ error bar on this measurement ($\sigma_{\hat{\rm A}^2}$) and the significance of the detection (defined as $\hat{\rm A}^2/\sigma_{\hat{\rm A}^2}$). As our statistic, we chose to use $\hat{\rm A}^2$ because it provides an unbiased estimator for the power in the timing residuals that can be attributed to a GWB. All our modifications are available in the most recent release of the \textsc{detectGWB} plugin.

We applied our method to all realizations of all our simulations. In
Figure~\ref{Fig:all_hd} we show the covariances obtained as a function
of the angle between each pulsar pair. In the left-hand column we show
an individual realization for each noise process.  In the right-hand
column we show the covariances after averaging over the 1000
realizations.  In the next section we discuss each of these noise
processes in detail.

\section{Results and discussion}\label{Sec:res_disc}

We calculated the distribution of $\hat{\rm A}^2$ values from each of the 1000 realizations of every simulated noise process.  In
Table~\ref{Tab:A2histo} we list the corresponding mean and standard deviation of the distribution. For the simulations that do not include a GWB, we also determine the $\hat{\rm A}^2$ value that is exceeded in only 5\% of the simulations.  This value (``DT5'') represents the detection threshold that would have a 5\% false alarm probability. We also determine how many realizations of S$_{\rm gwb}$ exceed this threshold (P$_{\rm DT5}$), i.e. the probability of detection at this threshold. In other words, DT5 represents the correct detection threshold for the analyzed type of noise, while P$_{\rm DT5}$ is the probability of detection if the GWB amplitude is 1$\times 10^{-15}$ and we assume DT5 as the actual detection threshold. We trial various mitigation methods to reduce DT5 and thus increase P$_{\rm DT5}$. These mitigation methods are described in detail below and the corresponding results are also listed in Table~\ref{Tab:A2histo}.  For these results we apply the same mitigation routine to S$_{\rm gwb}$ as we have applied to the simulation being processed.

\begin{table*}
 \centering
 \small
 \caption{Summary of the simulation results. In the columns are
   indicated, respectively, the simulation name, the applied
   mitigation routine, mean $\hat{\rm A}^2$, standard deviation of the $\hat{\rm A}^2$ distribution, DT5 and
   P$_{\mathrm{DT5}}$. The mean, standard deviation and DT5 values are given with respect to 10$^{-30}$}\label{Tab:A2histo}
 \begin{tabular}{c l c c c c}
 \hline
 \textbf{Simulation} & \textbf{Mitigation routine} & \textbf{Mean} & \textbf{Standard deviation} & \textbf{DT5} & \textbf{P$\rm _{DT5}$} \\
 \hline

 S$_{\rm gwb}$   & $-$                                & 1.2  & 0.53        & $-$ & $-$         \\                
               & Clock: offset subtraction        &   1.2  & 0.53        & $-$ & $-$         \\                
               & Clock: \citealt{hc12}        &   0.98  & 0.43        & $-$ & $-$         \\                
               & Ephemeris: cosine subtraction    &   1.2  & 0.53        & $-$ & $-$         \\                
               & Ephemeris: \citealt{dh13}     &   0.40  & 0.24        & $-$ & $-$         \\                
               & Ephemeris: \citealt{ch10} &   0.88  & 0.40        & $-$ & $-$         \\ 
               
S$_{\rm utn}$    & $-$                                & 0.0050 & 0.18        & 0.32        & 98 \%         \\

S$_{\rm clk}$    & $-$                                & 0.68  & 0.36        & 1.3        & 33 \%         \\
               & Clock: offset subtraction        &   $-$0.0066  & 0.14        & 0.25         & 99 \%         \\
               & Clock: \citealt{hc12}         &   0.0043  & 0.052        & 0.089        & 99 \%         \\
               
 S$_{\rm eph}$   & $-$                                &    0.36 & 0.19        & 0.74        & 81 \%         \\
               & Ephemeris: cosine subtraction    &  $-$0.00019  & 0.12        & 0.20        & 99 \%         \\
               & Ephemeris: \citealt{dh13}  & $-$0.021  & 0.034        & 0.041        & 97 \%         \\
               & Ephemeris: \citealt{ch10}  &  0.090  & 0.13        & 0.31        & 95 \%         \\
               
 S$_{\rm{tt}}$ & $-$ & 4.9  & 0.33        & 5.5        & 0 \%         \\
               & Clock: offset subtraction &$-$0.00067  & 0.31        & 0.0053       & 93 \%         \\
               & Clock: \citealt{hc12}  & 0.0040  & 0.053        & 0.097        & 99 \%         \\
               
 S$_{\rm{de}}$ & $-$ &  0.44   & 0.14        & 0.68        & 86 \%         \\
               & Ephemeris: cosine subtraction & 0.0029  & 0.14        & 0.25        & 99 \%         \\
               &  Ephemeris: \citealt{dh13} & $-$0.024  & 0.12        & 0.18        & 83 \%         \\
               & Ephemeris: \citealt{ch10}& 0.025  & 0.13        & 0.25        & 97 \%         \\
 S$_{\rm{ie}}$ (20\,ns)  & $-$ & $-$0.0021  & 0.069        & 0.11        & 99 \%       \\ 
 S$_{\rm{ie}}$ (200\,ns)  & $-$ & $-$0.36  & 2.1        & 3.1        & 0 \%        \\
 S$_{\rm{sw}}$   & $-$  & 1.6  & 0.14        & 1.8      & 11 \%         \\
               & Solar wind: \textsc{tempo2} model&  0.43  & 0.12        & 0.64        &  $-$        \\
               & Solar wind: adaptation of \citealt{kcs13} & 1.2  & 0.061        & 1.3        & $-$         \\
 \hline
 \end{tabular}
\end{table*}

\subsection{GWB and uncorrelated timing noise}

\begin{figure}
 \centering
 \includegraphics[scale=0.45,trim=0.5cm 0 0 0]{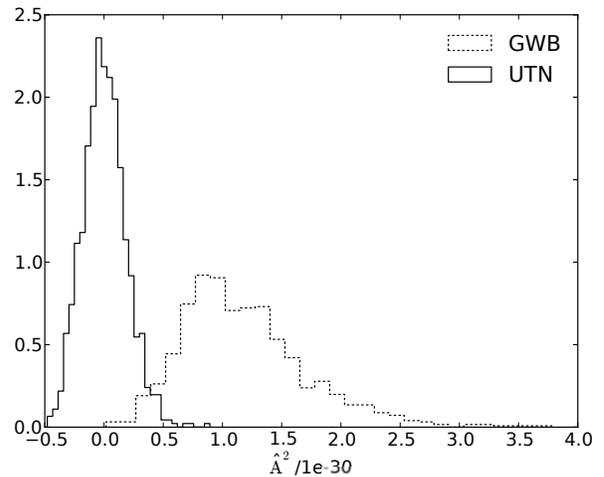}
 \caption{Normalized $\hat{\rm A}^2$ distributions obtained for simulations
   S$_{\rm utn}$ (solid) and S$_{\rm gwb}$
   (dotted).}\label{Fig:a2_rn+gwb}
\end{figure}

Figure~\ref{Fig:a2_rn+gwb} shows the $\hat{\rm A}^2$ distribution obtained for S$_{\rm gwb}$ (dotted histogram).  The mean
value, 1.2$\times 10^{-30}$, is slightly higher than the expected
(1$\times 10^{-30}$).  This is a known bias within the Y11 algorithm
that occurs because the individual covariances are not
independent. Y11 accounted for this bias by using simulations to
determine a scaling factor.  For our work this bias is small and does
not affect our conclusions.  The top panel in Figure~\ref{Fig:all_hd}
shows that we successfully recover the expected Hellings \& Downs
signature.

The solid-line histogram in Figure~\ref{Fig:a2_rn+gwb} shows the results
from S$_{\rm utn}$ (uncorrelated noise process).  As expected, the
mean of this histogram (5$\times 10^{-33}$) and the corresponding
angular covariances are statistically consistent with zero. The
P$_{\rm DT5}$ of this simulation indicates that the $\hat{\rm A}^2$ values
obtained from 98\% of the S$_{\rm gwb}$ simulations are above the detection threshold. It is therefore unlikely that, with
a careful analysis of the angular correlations in Figure~\ref{Fig:all_hd} and the false alarm probability, the S$_{\rm utn}$
simulations could be misidentified as a GWB.

\subsection{Clock-like error}\label{Sec:clock}

\begin{figure}
\includegraphics[scale=0.5,trim=.5cm 0 0 0]{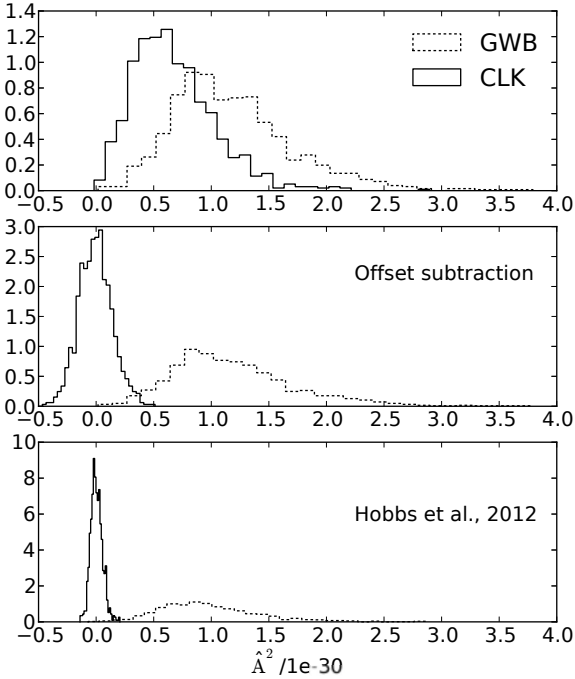}
\caption{In the upper panel are shown the normalized $\hat{\rm A}^2$
  distributions as computed for simulations S$_{\rm clk}$ (solid
  histograms) and S$_{\rm gwb}$ (dotted histogram). In the central and
  the bottom panels we show the same distributions, after the
  application of the two mitigation routines described in Section
  \ref{Sec:clock}.}\label{Fig:a2dealing_clk}
\end{figure}

The top panel of Figure~\ref{Fig:a2dealing_clk} contains the $\hat{\rm A}^2$ distribution obtained for S$_{\rm clk}$, overlaid with the $\hat{\rm A}^2$ values computed for S$_{\rm gwb}$. The two distributions significantly
overlap. The Y11 algorithm produces non-zero $\hat{\rm A}^2$ values with a mean
of 6.8$\times 10^{-31}$ and a DT5 of 1.3$\times 10^{-30}$. The P$_{\rm
  DT5}$ is 33\%.  Blind use of a detection code can therefore lead to
a false detection of a GWB in the presence of such a clock signal.  However, the
angular correlations for a clock error as a function of angular
separation take the form of a constant offset (see
Figure~\ref{Sec:ang_correlations}) and are easily distinguishable by eye from the
Hellings \& Downs curve. We can therefore update the GWB detection code to account for the possibility
of clock errors.

We consider two general approaches: 1) we can mitigate the effects
that the spurious signal has on the angular covariances or 2) we can
identify and subtract the signal directly from the time series. In the
first method, we note that the clock errors lead to an offset in
the angular correlations. We can therefore attribute any offset to
clock errors and update the detection algorithm to search for the
Hellings \& Downs curve with a mean removed. In the second method, we
can first measure the clock errors from the timing data, subtract them from the residuals
and then apply the detection algorithm.

For the first case, we update the Y11 algorithm to include a constant
offset, $\gamma$, when fitting for the $\hat{\rm A}^2$ values. i.e., we fit the
following function to our measured covariances:

\begin{equation}
f(\theta_{ij}) = \hat{\rm A}^2  C(\theta_{ij}) + \gamma.
\end{equation}\label{Eq:hd+const}
The resulting $\hat{\rm A}^2$ distribution for S$_{\rm clk}$ and S$_{\rm gwb}$
after this modification is shown in the central panel of
Figure~\ref{Fig:a2dealing_clk} (respectively the solid and dotted
histogram). The mean of the S$_{\rm clk}$ distribution is now
statistically consistent with zero. The standard deviation of the
distribution is reduced with respect to the non-mitigated results by a
factor of $\sim2.5$ and the DT5 is reduced by a factor of $\sim5$. The application of this correction
routine does not affect S$_{\rm gwb}$.

In the second case, we determine and remove the clock signal before
applying the GWB search algorithm \citep{hc12}. This procedure
requires to search for a common signal in the time series, and it is carried out by simultaneously fitting all the timing residuals with an even grid of linearly interpolated values. We choose a regular sampling of 150\,days, but note that the results
do not significantly depend upon this choice\footnote{We confirmed
  this by trialling grid spacings of 100\,days and 200\,days by iterating the simulations 100 times instead of 1000. The mean and standard deviation of the $\hat{\rm A}^2$ distributions for S$_{\rm gwb}$ are, in the case of a grid
  spacing of 100\,days, 9.9$\times 10^{-31}$ and 4.4$\times
  10^{-31}$. In the case of a grid spacing of 200\,days, they result
  1.2$\times 10^{-30}$ and 5.4$\times 10^{-31}$. In the case of
  S$_{\rm clk}$, means and standard deviations of the A$^2$
  distributions are 5.8$\times 10^{-33}$ and 4.9$\times 10^{-32}$ for
  a 100\,days grid spacing, and 4.4$\times 10^{-33}$ and 7.2$\times
  10^{-32}$ for a 200\,day grid spacing.}.

As the mean of the Hellings \& Downs curve is non-zero, the clock
subtraction procedure will identify it as a clock signal and remove
it.  Therefore, the fit for $\hat{\rm A}^2$ performed by \textsc{detectGWB}
needs to be updated to fit the measured covariances with the following
function:

\begin{equation}
 g(\theta_{ij}) = \hat{\rm A}^2 ( C(\theta_{ij}) - \langle C \rangle)
\end{equation}\label{Eq:hd-mean}
where $\langle C \rangle$ is the mean value of the Hellings \& Downs
curve computed on the angular separations determined by the pulsars in
the sample.  Our resulting $\hat{\rm A}^2$ distributions are shown in the
bottom panel of Figure~\ref{Fig:a2dealing_clk} for S$_{\rm clk}$ and
S$_{\rm gwb}$.  We obtain a distribution mean
that is consistent with zero for S$_{\rm clk}$. The standard deviation
and the DT5 decrease with respect to the non-mitigated simulations
by factors of $\sim7$ and $\sim15$ respectively. Applying
this correction routine to S$_{\rm gwb}$, the mean of the $\hat{\rm A}^2$
distribution decreases by about 18\%. It is statistically consistent
with the average $\hat{\rm A}^2$ of the non mitigated simulations (see
Table~\ref{Tab:A2histo}).

We therefore conclude that:

\begin{itemize}
\item Without correction, clock errors may yield large values of $\hat{\rm A}^2$
  when searching for a GWB in PTA data sets, and hence false
  detections;
\item It is possible to correct for clock errors without affecting the
  sensitivity to a GWB signal;
\item To measure and remove the clock signal from the time series
  before searching for the GWB is a satisfactory mitigation method.
\end{itemize}

\subsection{Ephemeris-like error}\label{Sec:eph}

\begin{figure}
\includegraphics[scale=0.5,trim=.5cm 0 0 1.2cm]{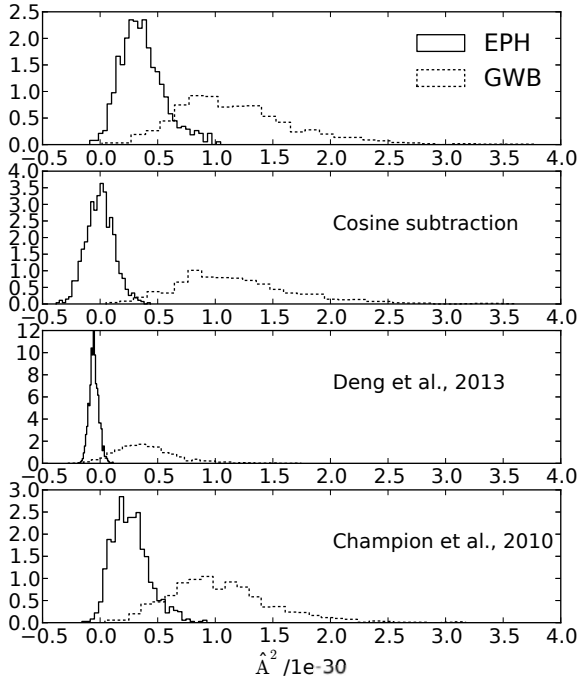}
\caption{In the upper panel are shown the normalized $\hat{\rm A}^2$
  distributions as computed for simulations S$_{\rm eph}$ (solid
  histograms) and S$_{\rm gwb}$ (dotted histogram). In the central and
  bottom panels we show the same distributions, after the application
  of the two mitigation routines described in this Section
  (\ref{Sec:eph}).}\label{Fig:a2dealing_eph}
\end{figure}

The $\hat{\rm A}^2$ values obtained for S$_{\rm eph}$, without applying any
of the clock correction routines discussed above, are shown in the top
panel of Figure~\ref{Fig:a2dealing_eph} (solid histogram) overlaid on
the GWB values.  The histograms partially overlap although the angular
covariances shown in Figure~\ref{Fig:all_hd} are clearly different.

As for the clock-like errors, we can 1) update the detection code to
account for the signature left by ephemeris errors on the angular
covariances or 2) we can first attempt to measure and remove the
ephemeris errors from the pulsar time series.

In the first case, we update the Y11 detection algorithm to include a
cosinusoidal function (see Appendix~\ref{App:a1}) while fitting for
$\hat{\rm A}^2$, i.e., we fit the following function to the measured
covariances:
\begin{equation}
f(\theta_{ij}) = \hat{\rm A}^2 C(\theta_{ij}) + \alpha \cos{\theta_{ij}}.         
\end{equation}
The resulting $\hat{\rm A}^2$ values are shown in the second panel of
Figure~\ref{Fig:a2dealing_eph}, both for S$_{\rm eph}$ and S$_{\rm
  gwb}$. The distribution mean for S$_{\rm eph}$ is now statistically
consistent with zero and the standard deviation is reduced by about a
factor 2. The S$_{\rm gwb}$ distribution is unaffected.

There are two published methods for measuring and removing the
ephemeris errors from the time series.  The first is a generalization
of the \citet{hc12} method for determining the clock errors and is
similar to that described by \citet{dh13}.  In this process, we
simultaneously fit for the three components of ${\bf e}(t)$ in
Equation~\ref{Eq:res_pe} at a uniform grid of epochs. These parameters
can subsequently be included in the pulsar timing models and, hence,
subtracted from the residuals.  The second method was proposed and
used by \citet{ch10}.  This method measures errors in the masses of
known solar-system planets. For the S$_{\rm eph}$ data set we did not
simulate errors in particular planetary masses and so do not expect
that this technique will reduce the timing residuals. However, we
apply this method to the data to consider whether it will affect the
detectability of a GWB. When applying these correction routines, a
cosinusoidal function has been removed from the angular
covariances. Therefore, we update the Y11 algorithm to fit the angular
covariances for the following function:
\begin{equation}
 g(\theta_{ij}) = \hat{\rm A}^2 ( C(\theta_{ij}) - \beta \cos \theta_{ij})
\end{equation}\label{Eq:hd-cosine}
where $\beta$ is given by a preliminary fit of a cosine function to
the theoretical Hellings \& Downs curve sampled at the angular
separations of the pulsars in the sample.

Our results are shown in the two bottom panels of
Figure~\ref{Fig:a2dealing_eph}.  The \citet{dh13} method correctly
removes the ephemeris errors and the mean of the resulting histogram
is consistent with zero.  However, it has also absorbed
much of the GWB: the mean of the GWB histogram has decreased by a
factor 3. As expected, the \citet{ch10} method is not as effective as the \citet{dh13} algorithm in identifying and removing the ephemeris contribution. When applied to S$_{\rm gwb}$, it decreases the distribution
mean by a factor 1.36, leaving it statistically consistent with the
unaltered value.

We therefore conclude that:

\begin{itemize}
\item Without correction, planetary ephemeris errors may yield
  sufficiently large $\hat{\rm A}^2$ values to lead to false detections when searching
  for a GWB in PTA data sets;
\item Fitting a cosine to the angular covariances, as well as the
  Hellings \& Downs curve, removes the bias resulting from
  random ephemeris errors but does not reduce the noise in the A$^2$ estimator: the standard deviation is not significantly
  reduced;
\item Using the \citet{dh13} approach does correct the ephemeris
  errors, but also absorbs significant power from the GWB signal;
\item If the error in the ephemeris was known to be solely caused by
  errors in planet masses then one can use the \citet{ch10} method,
  which does not significantly affect the GWB search.
\end{itemize}

In these sections we simulated red noise processes that have
an identical power spectrum as a GWB with $A = 10^{-15}$.  Hereafter,
we consider other noise processes that may be present in the data, but
have different spectral shapes.

\subsection{Realistic clock and ephemeris error levels}

It is challenging to determine the actual amount of noise from the
clock or ephemeris errors in real PTA data. We can consider the signals injected in S$_{\rm
  tt}$ and S$_{\rm de}$ to be upper estimates of the real noise processes
given by clock and ephemeris errors. Nevertheless, we are aware that
the real errors might actually be higher. This could be caused by, e.g., still
unknown systematics in the analysis algorithms or, in the case of the
ephemeris, the occurrence of unknown bodies in the solar system or the
influence of asteroids. For instance \citet{hh10} compare three
different ephemerides and note that the JPL ephemeris does not include
a ring of asteroids that can shift the position of the barycentre by
$\sim 100$\,m. The results shown in Table~\ref{Tab:A2histo} indicate
that none of the $\hat{\rm A}^2$ distributions obtained for S$_{\rm tt}$ and
S$_{\rm de}$ are consistent with zero.

\begin{figure}
\centering 
\includegraphics[width=6cm,angle=-90]{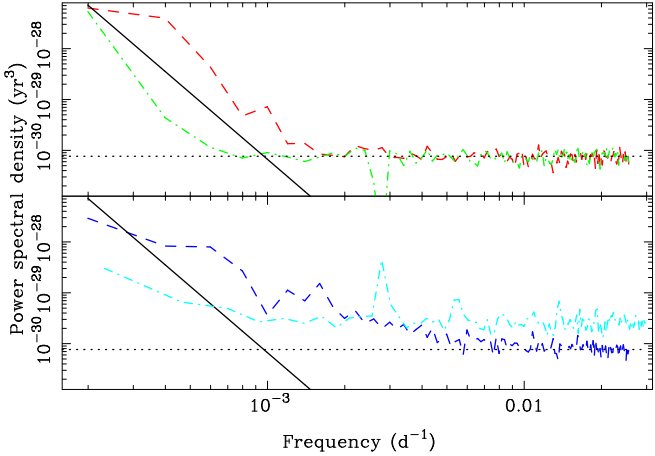}
\caption{Power spectra averaged over the 20 simulated pulsars for
  S$_{\rm tt}$ (dashed line) and S$_{\rm de}$ (dot-dashed line) are shown in the top panel. S$_{\rm ie}$ at
   200\,ns (dashed line) and S$_{\rm sw}$ (dot-dashed line) are shown in the bottom panel. In both panels the expected
  power spectral density for a GWB with A = 10$^{-15}$ is shown as the black solid line. The power spectral density corresponding to 100\,ns of white noise is indicated as the black dotted line.}\label{Fig:spectra}
\end{figure}

In the top panel of Figure~\ref{Fig:spectra} we show the power
spectra for S$_{\rm tt}$ and S$_{\rm de}$, averaged over the 20
pulsars simulated for one realization, along with the theoretical
power spectrum of a GWB with A = 10$^{-15}$. The power spectrum of
S$_{\rm tt}$ is higher than the GWB's, roughly corresponding to a
background amplitude of 2.1$\times$10$^{-15}$. The approximate
equivalent amplitude of S$_{\rm de}$ is lower, about
7.3$\times$10$^{-16}$. However, the slopes of the two spectra are very
similar to the predictions for a GWB.

\begin{figure}
\centering
\includegraphics[scale=0.48,trim=4cm 6cm 4cm 7cm]{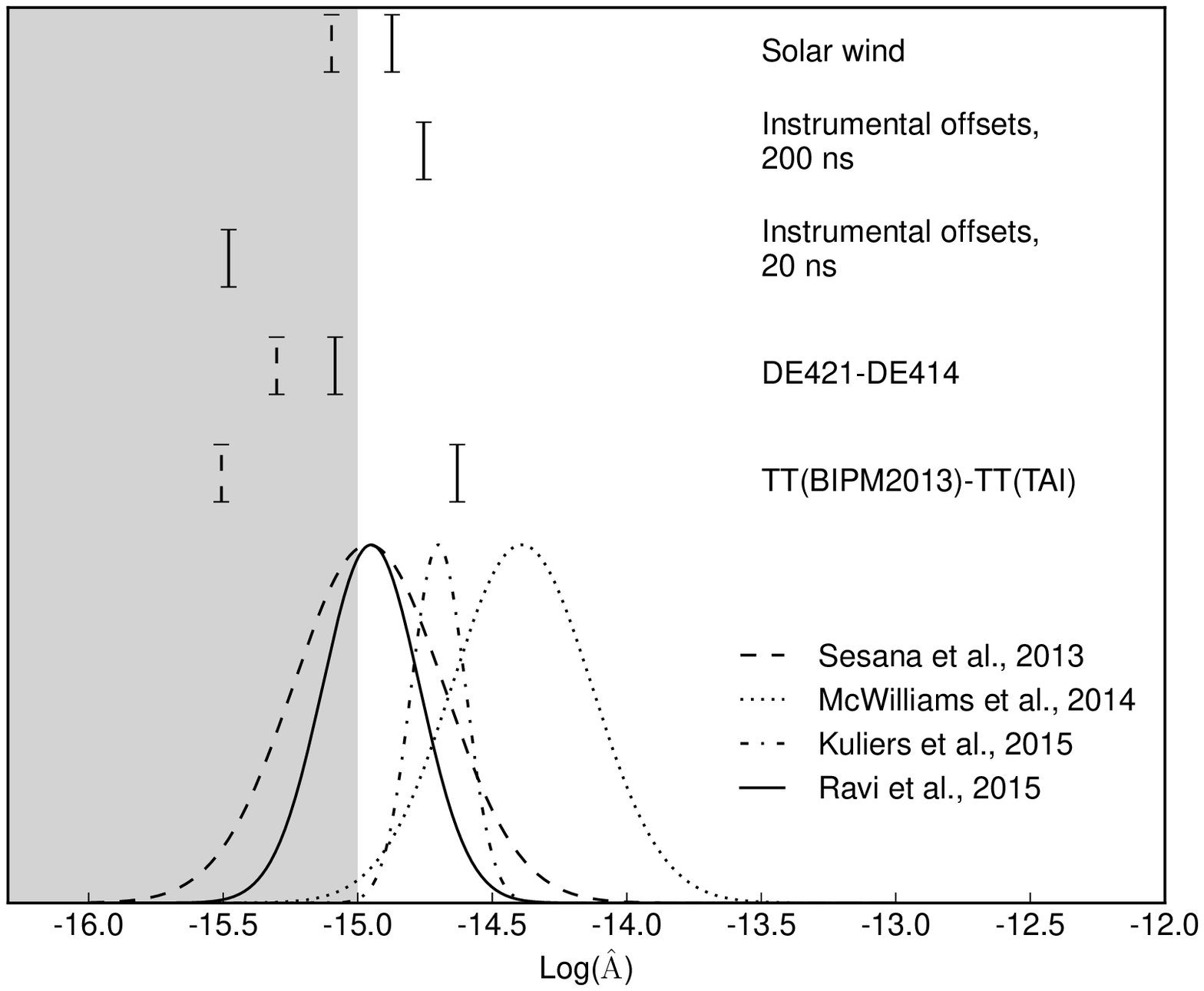}\\
\caption{In the lower part of the diagram we show the GWB amplitude
  ranges expected by four recent GWB models (fourth model of
  \citealt{ses13b}, \citealt{mo14}, \citealt{ko15} and
  \citealt{rw15}. The shaded region designates the latest 95\% upper
  bound to A$^2$ (Shannon et al., submitted). In the upper part of the
  Figure, we show the 95\% $\hat{\rm A}^2$ levels as obtained for S$_{\rm tt}$,
  S$_{\rm de}$, S$_{\rm ie}$ (at 20 and 200\,ns) and S$_{\rm sw}$. The
  solid and dashed (if present) bars indicate, respectively, the
  95\% bound without and with (dashed) mitigation algorithm (the
  \citealt{hc12} for the clock errors, the \citet{ch10} for the
  ephemeris errors and the default \textsc{tempo2} model for the solar
  wind effects.}\label{Fig:a2levels}
\end{figure}

In Figure~\ref{Fig:a2levels} we show DT5 obtained for the
S$_{\rm tt}$ and S$_{\rm de}$ simulations without (solid bars) and with
(dashed bars) mitigation methods being applied (the Hobbs et al. 2012\nocite{hc12} and the Champion et al. 2010
\nocite{ch10} methods respectively). We also show the ranges of the GWB amplitude as predicted
by four recent models (Gaussian functions shown at the bottom of the figure). The shaded region designates the 95\% upper limit on
A$^2$ (Shannon et al., submitted).

For all the simulated effects different from a GWB, the range of the $\hat{\rm A}$ estimators show a non-negligible degree of overlap with the GWB amplitude interval predicted by some of the most recent models. In particular, this plot indicates that the clock
errors of S$_{\rm tt}$ are likely to be the most significant source of
false detections. However, as shown in Section~\ref{Sec:clock}, clock
errors can be mitigated using one of the algorithms that we
described. Errors in the ephemeris are at a smaller amplitude. They
can also lead to significant false detections if the actual GWB amplitude
is lower than $\sim 10^{-15}$ (as the current upper limit on the GWB amplitude
indicates). All the mitigation methods that we have considered for
planetary ephemeris errors have problems.  For instance, the
\citet{ch10} can only be applied to errors in the masses of known
objects.  \citet{ch10} suggested
that the only likely sources of error are in the masses of Jupiter
and/or Saturn.  With recent spacecraft flybys these errors are likely
to be small, and can easily be corrected using the \citet{ch10}
procedure without affecting any underlying GWB.  The method could, if
necessary, also be updated to account for the poorly known mass of any
asteroid that is expected to produce a significant residual.  Previously unknown
objects are more easy to deal with by using the
\citet{dh13} method.  However, that method currently absorbs a
significant part of the GWB signal. We are therefore studying whether
that method can be modified so that it does not significantly affect
any GWB signal, by, for example, ensuring that all errors in the
ephemeris are caused by orbiting bodies with a prograde or retrograde
motion, or constraining the errors to be close to the ecliptic plane.

We outline below an illustrative procedure to confirm that a GWB detection
is not caused by clock and ephemeris errors:

\begin{itemize}
\item It is always valid to remove a constant
  offset and a cosine function from the covariances before (or while)
  fitting the Hellings \& Downs curve.  This leads to a small
  reduction in the amplitude of the GWB, but that can be subsequently
  calibrated and it eliminates the possibility of a false alarm caused
  by clock or certain types of ephemeris errors.
\item If the constant offset measurement is statistically significant
  then we recommend first measuring the clock errors (using e.g., the
  Hobbs et al., 2012 method), removing them and then searching again for the
  GWB accounting for the subtracted signal.
\item If the cosine term is statistically significant then it is
  likely that an ephemeris error is present.  Mitigating such errors
  will depend upon their nature.  We can apply the \citet{dh13}
  procedure, reducing its impact on the GWB search by rotating the
  resulting vector components into the ecliptic plane. If the error is
  constrained to the ecliptic plane, it would be possible to fit only
  for two components of $\mathbf e$.  The power spectra of these
  components may also indicate excess power at the period of a known
  planet.  If so, one can apply the \citet{ch10} method.  A final
  search could be carried out to determine whether the errors
  corresponded to prograde or retrograde orbits. If true, then a fit
  to such errors can be carried out, thereby further reducing the
  noise.
\item If the GWB detection is still significant after the above
  procedure, then it will be necessary to determine the false alarm
  probability of the measured $\hat{\rm A}^2$ value and we note that, if the GWB amplitude is
at the low end of the ranges ranges predicted by the models, then the
signals we simulated can induce false detections even after being
``mitigated".
\end{itemize}

\subsection{Other correlated errors}

\subsubsection{Instrumental errors}

For S$_{\rm ie}$, the pulsar data sets are correlated - all
instrumental jumps occur at the same epoch. However, the measured
covariances (Figure~\ref{Fig:all_hd}) depend upon the size of the
offsets. An offset introduced by a receiver or processing system
upgrade might be up to a few microseconds. Individual cases can be
fitted and removed, but these removal procedures might leave residual
offsets up to tens or hundreds of nanoseconds.   The power spectral density for 200\,ns offsets is shown in the bottom panel of Figure~\ref{Fig:spectra}.  The power spectral density for smaller offsets can be estimated by simply scaling the line shown in this Figure. For instance, offsets of 20\,ns would have a power spectral density that is 100 times smaller than that shown for the 200\,ns offsets.

As Table~\ref{Tab:A2histo} shows, in the case of S$_{\rm ie}$ at 20\,ns,
the mean of the resulting $\hat{\rm A}^2$ distribution is statistically
consistent with zero, and DT5 is zero: the distribution does not
overlap with the results from S$_{\rm gwb}$. However, with a residual
error size of 200\,ns, the standard deviation increases by a factor
$\sim30$. In this case, the $\hat{\rm A}^2$ values do overlap those from S$_{\rm
  gwb}$. The shape of the angular correlations suggests that these
signals are unlikely to be misidentified with a GWB. Nevertheless,
such uncertainties do exist in real time series, and only careful
inspection of the data can ensure that they do not interfere with the
GWB detection.

\subsubsection{Solar wind}

The power spectral density of the residuals induced by the solar wind is shown as the dot-dashed line in Figure~\ref{Fig:spectra}.  This power spectral has a complex form. The entire white noise level is higher than given by the simulated white noise level (i.e., the solar wind is inducing excess white noise). A clear excess in power is observed with a 1 year periodicity and there is an increase in power at low frequencies.

The solar-wind simulations S$_{\rm sw}$ require more pre-processing than the
others.  When the line-of-sight to these pulsars passes
close to the Sun, the effects of the solar wind can become extremely
large (Table~\ref{Tab:ppta_msp} shows that two of the pulsars,
PSRs~J1022+1001 and J1730$-$2304, have an ecliptic latitude of less than
one degree). In these conditions, PTA teams usually do not collect ToAs.

We chose a conservative approach, by excluding simulated observations
within 14 degrees from the Sun. As the routines used to simulate the effect of the solar wind (described in detail in \citealt{yhc+07a}) have a
degeneracy near to 180 degrees, we also excluded these observations.

Figure~\ref{Fig:all_hd} shows that pulsar pairs with small and large
angular distances are, respectively, correlated and
anti-correlated. Table~\ref{Tab:A2histo} shows that the $\hat{\rm A}^2$
distribution obtained for S$_{\rm sw}$ is narrower than the S$_{\rm
  gwb}$ by a factor $\sim4$, but the mean is slightly higher:
1.6$\times 10^{-30}$.

An attempt to mitigate the impact of the solar wind is implemented in \textsc{tempo2} and is applied by default.  This process simply assumes a constant, spherically symmetric model for the solar wind and predicts the residuals based on the angle between the line-of-sight to the pulsar and the Sun.  The electron density at 1~AU is set to 4~cm$^{-3}$ by default (an analogous model is available in the \textsc{tempo} software package, with an electron density of 10~cm$^{-3}$ at 1~AU).  As shown in Table~\ref{Tab:A2histo} the application of this method reduces the mean of the obtained $\hat{\rm A}^2$ distribution by a factor of $\sim$4.

Since the solar wind variability induces fluctuations in the DM values, it is also possible to attempt a mitigation by applying existing algorithms that model the DM variations. The method of \citet{kcs13} is the one that is currently adopted by the PPTA for the same purpose, and uses multi-frequency data in order to obtain a grid of time-dependent DM values, computed as averaged estimates over the duration of the grid steps. Simultaneously, the frequency-independent noise over the same grid steps is modeled in order to avoid misidentifying frequency-independent timing noise as DM variations.

We do not simulate multi-frequency data. However, we can adapt the \cite{kcs13} method by attributing every offset in the time series
to DM variations caused by the solar wind. This is possible, as we know the simulated data that we are dealing with in this section do not contain any other low-frequency noise besides the solar wind signal. We thus estimated and removed the DM fluctuations by using a grid step of 100\,days, independently for each pulsar. This method is not particularly effective. Although it strongly reduces the standard deviation of the A$^2$ distribution, it leaves the mean almost unaltered.  The most likely explanation for this result is that the variations caused by the solar wind are narrow and cuspy.  They are not well modeled by linear interpolation of DM variations on a 100\,day timescale.

The other PTA teams typically use different methods to assess the effects of DM fluctuations. The EPTA relies on a Bayesian analysis of the red noise processes, and assumes a 2-parameter power law to model the spectrum of the DM variations \citep{la13}. Nanograv includes the effect of DM fluctuations in time offsets that remain constant over a 15-day window \citep{df13}.

The impact of the solar wind may be minimized by observing at the highest feasible frequency. However, as it is difficult to remove it completely, a simulation of the solar wind effects should be included to estimate the false alarm probability of any GWB detection. Further work on this topic would be extremely valuable.

\subsection{The role of the IPTA}

We have shown that correlated noise processes present in PTA data, if sufficiently large, will affect the GWB search.  We have only made use of a single, frequentist-based detection method, but our results will also be applicable to any frequentist or Bayesian detection routine as all the PTA detection algorithms rely on the same basic principles (that of maximum likelihood). Moreover, even though we can develop mitigation routines to account for most of these noise processes, unexpected correlated signals will also exist in the data.

None of the non-GWB noise processes that we simulated led to angular
covariances that were identical to the Hellings \& Downs curve
(Figure~\ref{Fig:all_hd}).  This curve has positive correlations for
small angular separations and for separations close to 180 degrees and
negative correlations around 90 degrees.  A robust detection of the
GWB would therefore include strong evidence of this specific shape to
the angular covariances. The PPTA pulsars offer a good sampling of the
angular separations between $\sim$25 and $\sim$140 degrees (see
Figure~\ref{Fig:hdcurve}), but they poorly cover the closest and the
widest angular separations.  To make a robust detection of the GWB
even more challenging, the PPTA pulsars are non-homogeneous.  A few
pulsars are observed with significantly better timing precision than
other pulsars. 

In contrast, the IPTA contains a much larger sample of pulsars (the
current IPTA observes a sample of about 50 pulsars, although, as for the PPTA, the pulsars are non-homogeneous) covering a wider
range of angular separations (shown in black dots in Figure
\ref{Fig:hdcurve}).  It will therefore be significantly easier to
present a robust detection of the GWB with IPTA data than with PPTA
data alone.  The IPTA also provides other ways to reduce the chance of
false detections. For instance, multiple telescopes observe the same
pulsars: a comparison between data sets from different telescopes will
allow us to identify instrumental and calibration errors, along with
other issues such as errors in the time transfer from the local
observatory clocks. The increased number of observing frequency bands
will also enable the IPTA data to be corrected for dispersion measure and
possibly solar wind effects more precisely than they could by a single
timing array project.

 \section{Summary and Conclusions}\label{Sec:conclusions}

We have simulated various correlated noise processes that might affect
PTA pulsar data. We studied their effect on the GWB search by using
a PPTA detection code based on a frequentist approach. We also
tested mitigation routines for some of the simulated
signals. Our main conclusions are that:

\begin{itemize}
\item blind use of a detection code without mitigation routines can
  lead to false identifications of a GWB signal;
\item errors in the terrestrial time standard are expected to have the
  largest effect, but such errors can be mitigated without affecting
  any underlying GWB signal;
\item errors in the planetary ephemeris will become important if the
  GWB signal is significantly lower than the current upper bounds. It
  is not trivial to develop an effective mitigation routine that does
  not affect the underlying GWB signal;
\item the effect of instrumental errors scales with their amplitude. Small offsets are unlikely to cause false detections. Much larger offsets may do so;
\item the solar wind will yield false detections if not properly modeled;
\item other correlated signals may be present in a PTA data set and
  may affect a GWB search.
\end{itemize}

With care, a robust detection of the GWB can be made.  However,
determining all the physical phenomena that can affect the data is
challenging and the combined data sets of the IPTA will be necessary
to provide confident detections of the GWB.

\section*{Acknowledgements}

The Parkes radio telescope is part of the Australia Telescope National
Facility which is funded by the Commonwealth of Australia for
operation as a National Facility managed by CSIRO. CT is a recipient
of an Endeavour Research Fellowship from the Australian Department of
Education and thanks the Australian Government for the opportunity
given to her. GH is a recipient of a Future Fellowship from the
Australian Research Council. Wilcox Solar Observatory data used in
this study were obtained via the web site
{\url{http://wso.stanford.edu}} at \texttt{2015:06:11\char`_22:55:55}
PDT courtesy of J.T. Hoeksema. The Wilcox Solar Observatory is
currently supported by NASA.

\appendix
\appendixpageoff

\section{Signature of an error in the planetary ephemeris}\label{App:a1}

Errors in the terrestrial time standard lead to the same induced timing residuals for different pulsars.  The covariance between different pulsar pairs can therefore easily be determined. However, determining the covariances between pulsar pairs caused by an error in the planetary ephemeris is more challenging and we derive the expected covariance here.

The timing residuals of a planetary ephemeris error are given by Equation~\ref{Eq:res_pe}. The covariance $C(\theta_{ij})$ between the
residuals for two pulsars $i$ and $j$ induced by such an error is:

\begin{equation}
 C(\theta_{ij}) = \frac{1}{c^2} \langle ({\bf e}\cdotp {\bf \hat{k}}_{\rm i})   ({\bf e}\cdotp {\bf \hat{k}}_{\rm j}) \rangle
\end{equation}
where, as in Equation~\ref{Eq:res_pe}, $\bf e$ is the time-dependent error in position of the SSB with respect to the observatory, $\bf \hat{k}_{\rm i}$ and $\bf \hat{k}_{\rm j}$ are unit vectors pointed toward pulsars $i$ and $j$.  The ensemble averages are determined over each observation. We write:
\begin{eqnarray}
{\bf e} \cdotp {\bf \hat{k}}_{\rm i} &=& |{\bf e}|\cos{\alpha_i} \\
{\bf e} \cdotp {\bf \hat{k}}_{\rm j} &=& |{\bf e}|\cos{\alpha_j} 
\end{eqnarray}
and note that $\theta_{ij}$ has already been defined as the angle between the two pulsars.

Applying the cosine formula for a spherical triangle we have:
\begin{equation}
\label{eqn:a4}
 C(\theta_{ij}) = (1/c^2) \left(\langle |{\bf e}|^2 \cos{\theta_{ij}} \rangle - \langle |{\bf e}|^2
 \sin\alpha_i \sin \alpha_j \cos{\beta} \rangle \right)
\end{equation}
where $\beta$ is the angle subtended at e by $\hat{k}_{\rm i}$ and $\hat{k}_{\rm i}$ in the spherical triangle connecting the ends of the unit vectors.

If the error vector $\bf e$ is uncorrelated with the pulsar directions, the second term goes to zero and thus:

\begin{equation}
 C(\theta_{ij}) = \langle |{\bf e}|^2 \rangle \cos{\theta_{ij}} /c^2.
\end{equation}

In this case, the expected covariances will therefore have a cosinusoidal shape. Of course, a mass error in a specific planet leads to a deterministic error.  However, errors induced by a large number of small masses may have a more different form. We therefore note that the actual covariances measured will only show a clear cosinusoidal form if this term dominates the second term in Equation~\ref{eqn:a4}.

\bibliographystyle{mn2e}
\bibliography{journals,modrefs,psrrefs,crossrefs}

\bsp
\label{lastpage}
\end{document}